\documentclass[journal=jctcce,manuscript=article,layout=twocolumn]{achemso}

\usepackage[english]{babel}

\usepackage{amsmath}
\usepackage{graphicx}
\usepackage{natbib}

\usepackage[colorlinks=true, allcolors=blue]{hyperref}

\title{Validation of the Alchemical Transfer Method for the Estimation of Relative Binding Affinities of Molecular Series}
\author{Francesc Saban\'es Zariquiey}
\affiliation{Computational Science Laboratory, Universitat Pompeu Fabra, Barcelona Biomedical Research Park (PRBB), C Dr. Aiguader 88, 08003, Barcelona, Spain}
\author{Adri\`a P\'erez }
\affiliation{Acellera Labs, C Dr Trueta 183, 08005, Barcelona, Spain}
\author{Maciej Majewski}
\affiliation{Acellera Labs, C Dr Trueta 183, 08005, Barcelona, Spain}
\author{Emilio Gallicchio}
\email{egallicchio@brooklyn.cuny.edu}
\affiliation[NYC]{Department of Chemistry, Brooklyn College of the City University of New York;  PhD Program in Chemistry, Graduate Center of the City University of New York; PhD Program in Biochemistry, Graduate Center of the City University of New York, USA}
\author{Gianni De Fabritiis}
\email{gianni.defabritiis@upf.edu}
\affiliation{Computational Science Laboratory, Universitat Pompeu Fabra, Barcelona Biomedical Research Park (PRBB), C Dr. Aiguader 88, 08003, Barcelona, Spain}
\alsoaffiliation{Acellera, Devonshire House 582 Honeypot Lane Stanmore Middlesex,
HA7 1JS United Kingdom}
\alsoaffiliation{Instituci\'o Catalana de Recerca i Estudis Avan\c{c}ats (ICREA), Passeig Lluis Companys 23, 08010 Barcelona, Spain}

\begin{document}
\maketitle

\begin{abstract}
The accurate prediction of protein-ligand binding affinities is crucial for drug discovery. Alchemical free energy calculations have become a popular tool for this purpose. However, the accuracy and reliability of these methods can vary depending on the methodology. In this study, we evaluate the performance of a relative binding free energy protocol based on the alchemical transfer method (ATM), a novel approach based on a coordinate transformation that swaps the positions of two ligands. The results show that ATM matches the performance of more complex free energy perturbation (FEP) methods in terms of Pearson correlation, but with marginally higher mean absolute errors. This study shows that the ATM method is competitive compared to more traditional methods in speed and accuracy and offers the advantage of being applicable with any potential energy function.
\end{abstract}

\section{Introduction}

The ability to accurately predict the binding free energy of a ligand to a protein can provide crucial information for drug discovery, as it allows for the identification of compounds that have a higher likelihood of binding to a target. Alchemical free energy calculations have become the leading tools in this field. \cite{jorgensen2004many,abel2017advancing,armacost2020novel}  Free energy approaches are especially relevant in hit-to-lead and lead optimization stages of drug design while dealing with a series of similar ligands. Both commercial and free tools for free energy calculations have been developed over the past few years, with extensive use in both academia and the pharmaceutical industry.\cite{zou2019blinded,gapsys2015pmx,bieniek2021ties,wang2015accurate,kuhn2020assessment} 

One of the most common approaches to alchemical calculations is Free Energy Perturbation (FEP). This method involves many distinct equilibrium MD simulations for all states along a $\lambda$ coordinate that modifies a ligand A into a ligand B alchemically. Commonly these simulations are split between 12 or more $\lambda$-intermediates where the two ligands are interchanged.\cite{mey_best_2020} One of the most common methodologies that use this approach is Schrödinger's FEP+.\cite{wang2015accurate} Another way to approach alchemical calculations is via Thermodynamic Integration (TI), as in the Amber implementation.\cite{lee2020alchemical} The main difference with FEP is that TI calculates the free energy difference by integrating the derivative of the Hamiltonian with respect to the alchemical progress parameter $\lambda$. The pmx\cite{gapsys2015pmx} protocol implements a similar strategy based on non-equilibrium trajectories. Although different, FEP and TI share a few common traits such as the adoption of a double-decoupling process that obtains the relative binding free energy from the difference of the alchemical free energies from separate solution and receptor legs, and the requirement of softcore potentials to avoid clashes and instabilities\cite{beutler1994avoiding,zacharias1994separation,gapsys2012new}. Custom alchemical topologies and the need for multiple simulations of distinct systems (receptor complex and ligand in solvent) tend to require more user expertise. Furthermore, they are usually not suited for ligand pairs with different net charges\cite{chen2018accurate}, leading to potential issues with the treatment of long-range electrostatic interactions and artifacts in the free energy estimates, unless complex correction factors are introduced.\cite{rocklin2013calculating}

Recently, a novel approach to performing alchemical calculations has been proposed. The Alchemical Transfer Method (ATM) is a protocol for the estimation of relative binding free energies based on a coordinate transformation that swaps the positions of two ligands. The method performs the calculation in a single solvent box and, unlike double-decoupling free energy perturbation approaches,\cite{Gilson:Given:Bush:McCammon:97,wang2015accurate,lee2020alchemical} avoids the split of the binding free energy calculation into receptor and solvation legs. Furthermore, ATM does not require the implementation of softcore pair potentials. ATM is implemented in the free and open-source OpenMM\cite{eastman2017openmm} molecular simulation package, allowing a simple and easy route to large-scale automated deployments and flexibility to employ  any potential energy function. 
In spirit, ATM is similar to the separated topologies method \cite{rocklin2013separated} with the difference that the latter achieves the transfer by decoupling the first ligand while coupling the second by modifying the force field parameters. Whereas in ATM, the perturbation is implemented as a coordinate displacement that swaps the position of the two ligands. Recently, this approach was reintroduced to be used in GROMACS.\cite{baumann2023broadening}
ATM can handle both relative (ATM-RBFE) and absolute (ATM-ABFE) binding free energy calculations. In this work we focus on testing the accuracy and feasibility of the RBFE approach.

All the different free energy estimation methods have their pros and cons and can vary in their accuracy and reliability. Therefore it is important to rigorously evaluate their performance against large and diverse benchmark datasets.

In this work, we aim to evaluate the performance of ATM,\cite{wu2021alchemical,azimi2022relative} using the dataset of Wang \textit{et al.},\cite{wang2015accurate} one of the most popular benchmarks for evaluating relative binding free energy protocols. We use ATM to calculate the difference in binding free energies for 330 ligand pairs across 8 different protein systems. We also compared our results with state-of-the art methodologies such as FEP+\cite{wang2015accurate}, Amber\cite{lee2020alchemical} and pmx\cite{gapsys2020large}. We show that ATM, a methodology that requires less expertise and preparation than alternative protocols, performs as well as other existing tools and even better from a correlation point of view.

\section{Methods}

The aim of this study is to further expand the benchmarking of ATM to a series of targets tested in other similar methodologies and evaluate whether it can provide accurate and reliable estimates of relative binding free energies for these systems. To address this question, we conducted a computational study in which we applied ATM to the dataset of Wang \textit{et al.}.\cite{wang2015accurate} This benchmark includes eight targets relevant to pharmaceutical research (MCL-1, TYK2, MCL-1, JNK1, PTP1B, BACE, Thrombin and p38) with a total of 330 ligand pairs.

ATM is based on a displacement coordinate transformation that swaps the positions of two ligands, one of which is initially placed in the binding site of the receptor and the other into the solvent bulk.\cite{azimi2022relative} The potential energies of the system before and after the displacement are combined into a $\lambda$-dependent potential function, such that the system is progressively transformed from the state in which the first ligand is bound to the receptor and the second is in solution, to the reversed situation in which the second ligand is bound to the receptor and the first is not. ATM protocol does not require soft-core pair potentials or modifications of the energy routines of the molecular dynamics engine, and it does not require splitting the binding free energy calculation into receptor and solvation legs.  ATM is implemented as an OpenMM plugin.\cite{ATMMetaForce-OpenMM-plugin} Further details of the methodology can be found in previous work.\cite{gallicchio2015asynchronous,wu2021alchemical,azimi2022relative} 

\begin{figure}[h!]
\centering
\includegraphics[width=\columnwidth]{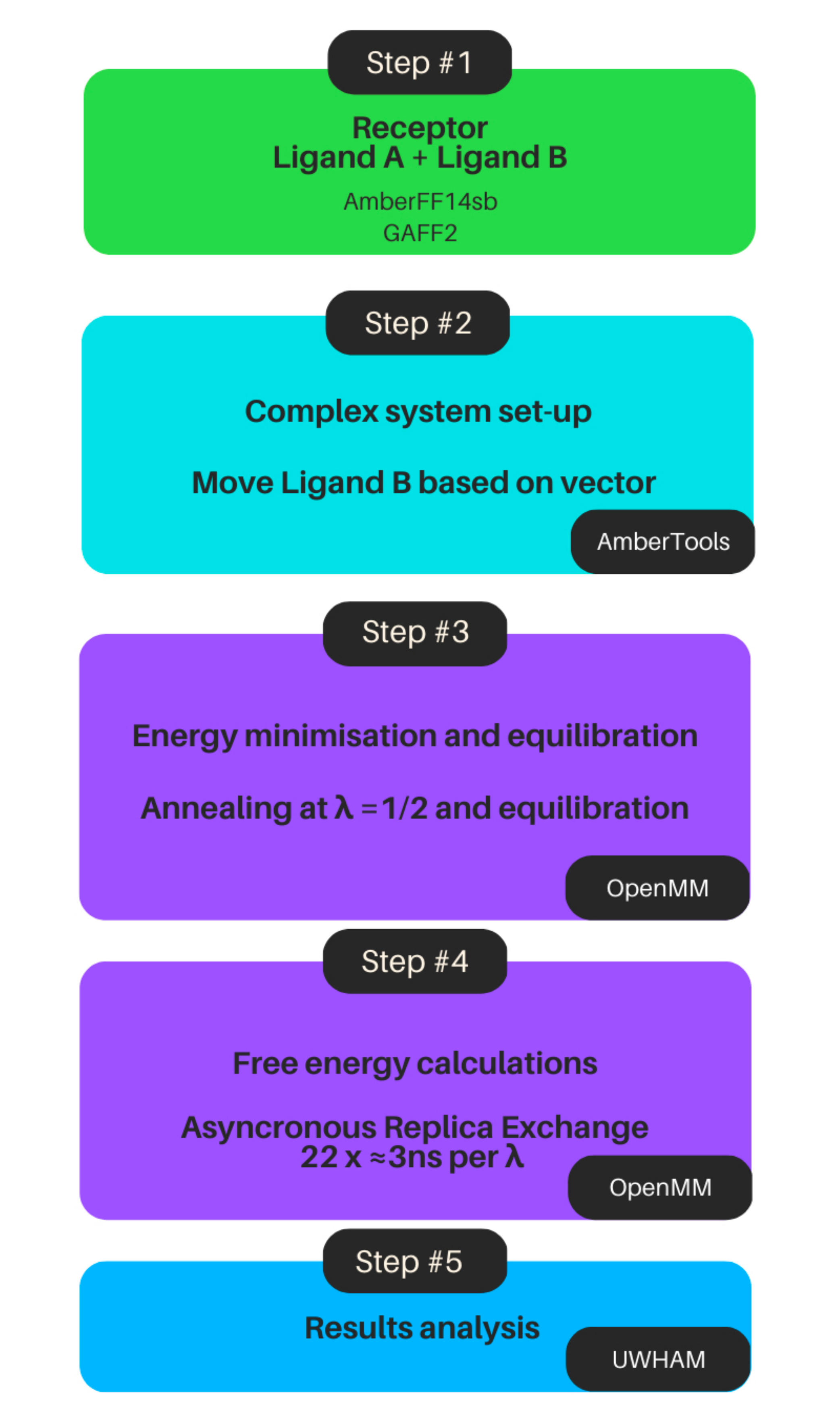}
\caption{ The AToM-OpenMM workflow used in this work. (1) Starting from the protein-ligand complexes from the Wang dataset\cite{wang2015accurate} the ligands are parametrised with GAFF2 and protein tropologies prepared in the Amber ff4SB forcefield. (2) System complexes are build with Ambertools and ligand B is displaced based on a vector. (3) Energy minimisation and equilibration is performed. Later an annealing and equilibration at $\lambda$=1/2 is performed. (4) Aysncronous replica exchange simulations are performed until a total sampling of at least 50ns is achieved. (5) After the simulations were finished, these were analyzed with the UWHAM package to obtain the calculated $\Delta\Delta G$ estimates.}
\label{fig:Workflow_ATM}
\end{figure}

We used the AToM-OpenMM package\citep{AToM-OpenMM} to set up and run the alchemical calculations. The AToM-OpenMM workflow (Fig. \ref{fig:Workflow_ATM}) prepares the complex systems for simulation using the LEaP program in AmberTools19.\cite{AmberTools19} Amber ff14SB parameters\cite{zou2019blinded,maier2015ff14sb} were assigned to the receptors while GAFF2/AM1-BCC\cite{wang2006automatic,he2020fast} were used for the ligands. Each complex system built in  LEaP consists of the receptor and a pair of aligned ligands. One of the ligands is selected to be translated along the diagonal of the solvent box so it is placed outside the receptor, ensuring at least three layers of water molecules in between. A restraining potential is also introduced to maintain geometrical  alignment between the two ligands aimed at enhancing the rate of convergence of the free energy estimate. The alignment restraints are based on the relative position and orientation  of the coordinate frames of the two ligands  defined by three chosen reference atoms.\cite{azimi2022relative} More information on the reference atoms selected for each system can be found in the Supporting Information (Supporting Figure \ref{fig:ref_alignment}).

Each complex system was solvated with a 10 \AA\ solvent buffer and with sufficient sodium and chloride ions to neutralize the system. The solvated complexes are minimized and thermalized at 300 K. Next, the system was annealed from the bound state ($\lambda$ = 0) to the symmetric alchemical intermediate ($\lambda$ = 0.5) for 250 ps. This step facilitates the creation of an initial configuration of the system without strong repulsive interactions at the alchemical intermediate state, which serves as the starting point for the subsequent Hamiltonian replica exchange\cite{gallicchio2015asynchronous} molecular dynamics that computes the free energies of the two ATM legs connecting the alchemical intermediate to physical end states at $\lambda=0$ and $\lambda=1$. C$\alpha$ atoms of the protein receptor were restrained using a flat-bottom harmonic restraining potential with a tolerance of 1.5 \AA.Additionally, we apply binding site restraints as a flat-bottom distance restraint between the geometrical centres of sets of receptor atoms that surround the binding site and ligand atoms. This defines the binding site volume as required by the quasi-chemical statistical mechanics formulation of molecular binding.\cite{Gilson:Given:Bush:McCammon:97} The restraints of the receptor and the ligands are optional and employed here to limit the conformational space that needs to be explored to reach convergence of the binding free energy estimate.
\begin{figure*}
\centering
\includegraphics[width=\linewidth]{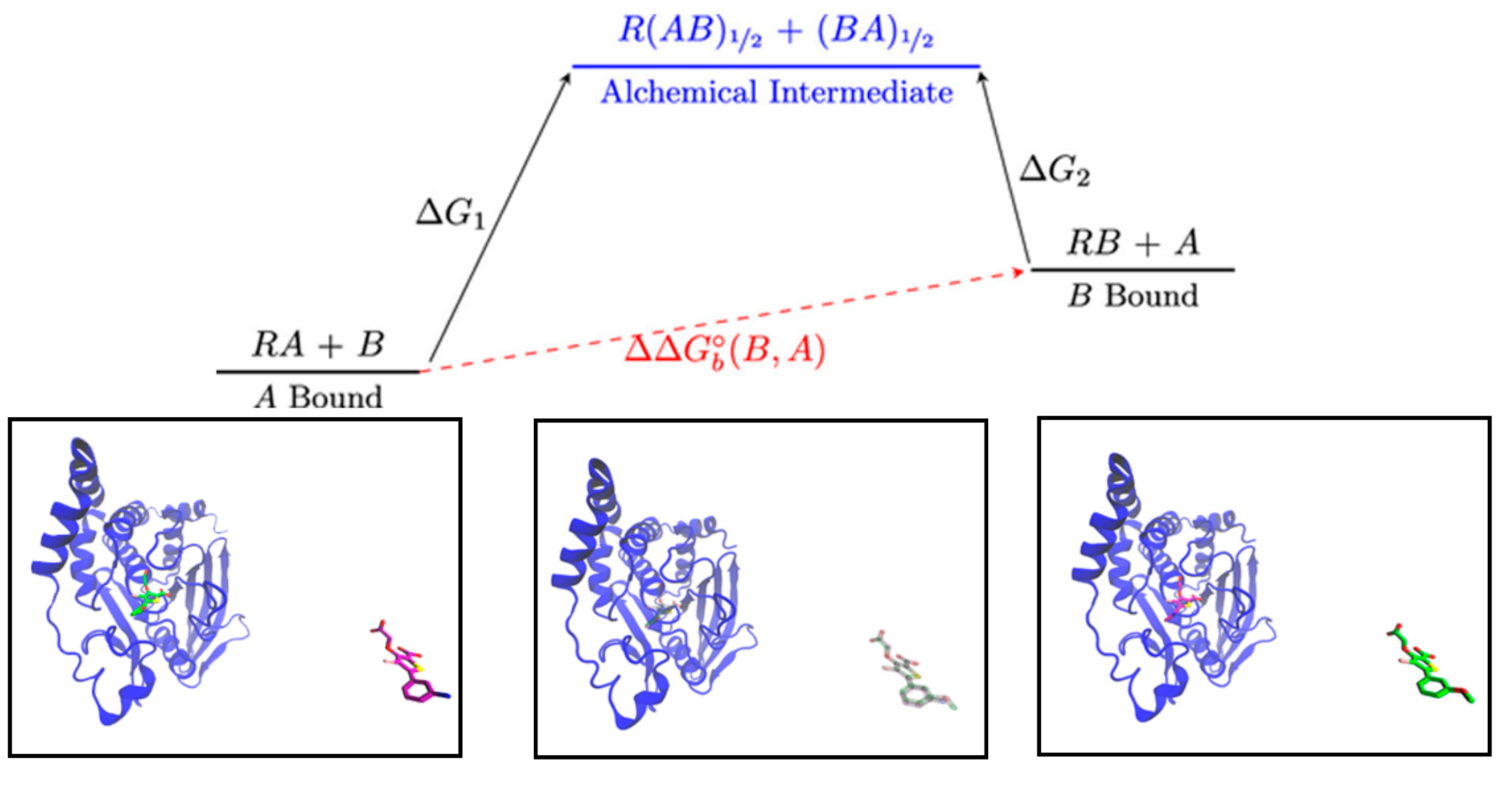}
\caption{ Top: Free energy diagram for an ATM-RBFE calculation, consisting of two independent legs connected to single alchemical intermediate state. The first leg starts at $\lambda$=0 where ligand A is bound to the receptor R's binding site, and ligand B is present in the solvent bulk. Leg 1 ends at $\lambda$=1/2 where both ligands A and B are simultaneously present at 50\% both in the binding site and in the solvent bulk. The second leg starts with ligand B bound to the binding site and ligand A in the solvent bulk and ends in the same alchemical intermediate. Bottom: Graphical representations for the described events.}
\label{fig:Scheme_ATM}
\end{figure*}

The softplus alchemical potential\cite{khuttan2021alchemical} was used for all calculations with 11 $\lambda$-states distributed between $\lambda$ = 0 and 0.5 for each of the two ATM legs (Supporting Table \ref{tab:params_table}). AToM-OpenMM performs asynchronous Hamiltonian replica exchanges in $\lambda$-space using the method described by Gallicchio \textit{et al.} \cite{gallicchio2015asynchronous} Exchanges were performed every 10 ps. To maintain a temperature of 300 K, a Langevin thermostat with a time constant of 2 ps was employed. Each ligand pair was simulated for a minimum of 50 ns per $\Delta\Delta G$ estimate. The sampling time has been chosen in order to be comparable to the aforementioned works, FEP+ studies sample between 36 and 60 ns per $\Delta\Delta G$ estimate whereas Lee \textit{et al.}\cite{lee2020alchemical} employed a total of 48 ns per ligand pair in Amber. In the case of pmx,\cite{gapsys2020large} calculations were carried out for 50 ns per pair for two force fields, GAFF and CGenFF. Since we performed ATM calculations on wall time rather than simulation time, as at the time of performing these calculations there was no support for simulation time-based runs, the sampled simulation time is similar but not identical for all ligand pairs.  Binding free energies and their corresponding uncertainties were calculated from the perturbation energy samples using the UWHAM method.\cite{tan2012theory} The obtained relative binding free energies ($\Delta \Delta G$) were compared to experimental measurements in terms of the mean absolute error (MAE), root mean square error (RMSE), and Pearson correlation coefficient. The obtained values are compared to the corresponding values reported in the literature.\cite{wang2015accurate,lee2020alchemical,gapsys2020large}

The parallel replica exchange alchemical molecular dynamics simulations were performed with the OpenMM 7.7\cite{eastman2017openmm} MD engine and the ATM Meta Force plugin\cite{ATMMetaForce-OpenMM-plugin} using the CUDA platform on NVIDIA RTX 2080 Ti cards.

\subsection{Results}

We conducted a comparison of ATM's relative binding free energies ($\Delta \Delta G$) estimates for the benchmarking dataset of Wang \textit{et al.}\cite{wang2015accurate} with those of commercial and open-source alchemical approaches: FEP+,\cite{wang2015accurate} Amber.\cite{lee2020alchemical} and pmx.\cite{gapsys2020large} In addition to differences in the methodology, these comparisons test variations in protein force fields, ligand parameterization techniques, and differences in the behavior of MD packages as potential sources of deviation in the obtained results. In this study, we have chosen to maintain the parameters described in the previous ATM publications as we believe it would provide a fair and consistent comparison.

\begin{figure}[h!]
\centering
\includegraphics[width=\columnwidth]{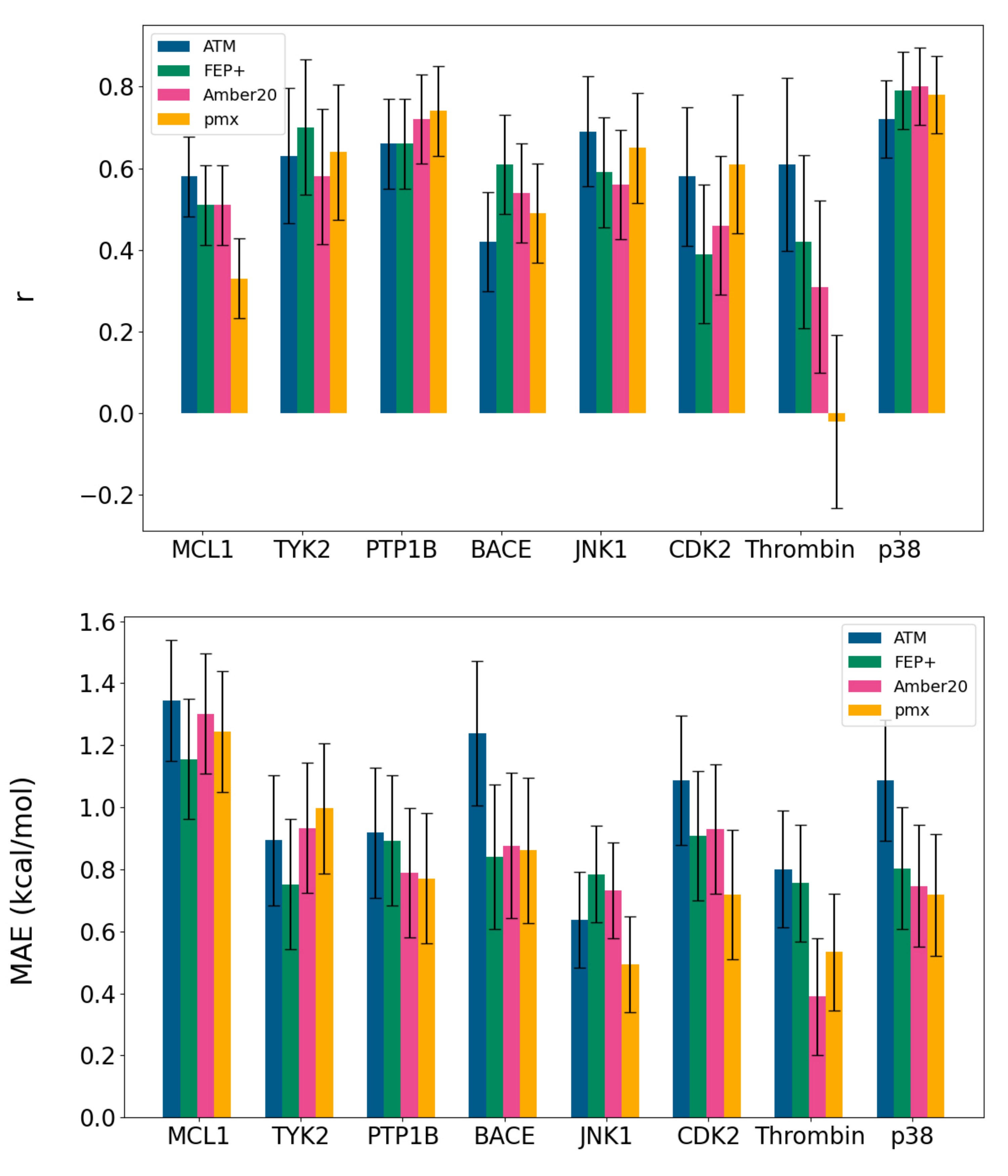}
\caption{ (Top) Pearson correlation (r) and (bottom) Mean Absolute Error (MAE) for each protein-ligand system calculated with ATM and reported estimates using the alternative methodologies FEP+\cite{wang2015accurate}, Amber\cite{lee2020alchemical} and pmx\cite{gapsys2015pmx}.}
\label{fig:comparison_corrs}
\end{figure}

\begin{figure*}
\centering
\includegraphics[width=\linewidth]{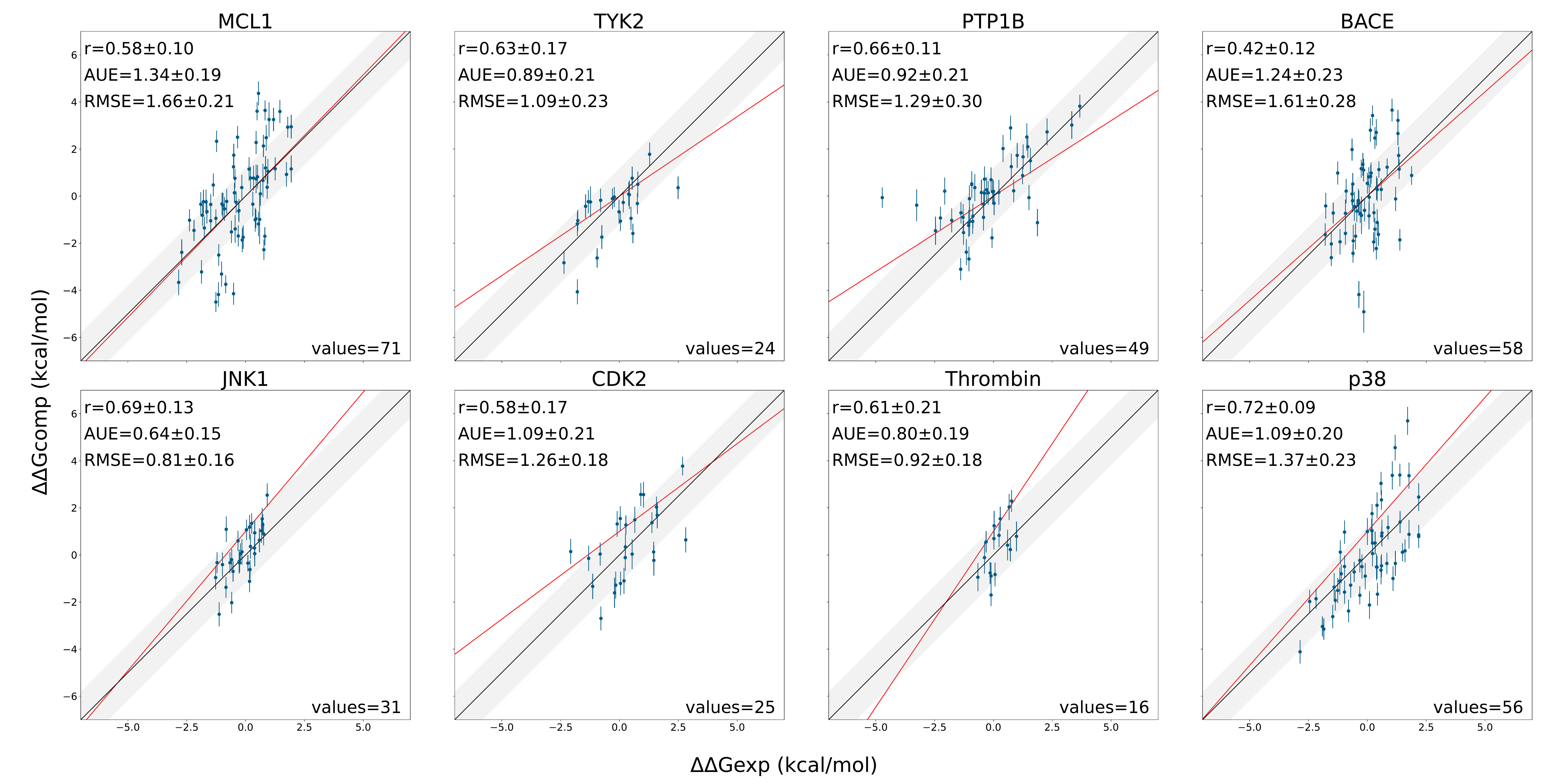}
\caption{ Performance of the Alchemical Transfer Method (ATM) for each protein-ligand system studied. The calculated $\Delta\Delta G$ estimates are plotted against their corresponding experimental values. MAE is in kcal/mol, $r$ is Pearson correlation and values refer to the number of ligand pairs evaluated for each system.}
\label{fig:prots_ATM}
\end{figure*}

The results of the simulations are displayed in Figures \ref{fig:comparison_corrs} and \ref{fig:prots_ATM}, which highlight the relative (Pearson correlation) and absolute (MAE) performance of the method. Table \ref{tab:result_table} contains a comparison against other free energy methods. Comparison against the other mentioned methodologies can be found in Supporting Figures \ref{fig:comparison_corrsSI_1}, \ref{fig:comparison_corrsSI_2} and \ref{fig:comparison_RMSE}. We observe that ATM performs similarly to the other approaches in overall Pearson correlation (0.59), with values for specific systems ranging from 0.42 to 0.71. ATM's Pearson correlation coefficients are particularly  good  for the MCL-1, JNK1, and Thrombin datasets, where it outperforms the other methods albeit only by relatively small margins. For the other protein targets, Pearson correlation metrics fall within those of the other methodologies. For instance, in the case of p38, the observed correlation is 0.71 for ATM, which is the lowest when compared to the other approaches. However, the difference is not significant as the results for all methods fall within the measurement error. Despite these positive observations, we did encounter some difficulties with the BACE dataset, as we obtained the lowest correlation (0.42) among the methods which is significantly lower than the best correlation value obtained using the FEP+ methodology (0.61). It is worth mentioning that the range of $\Delta\Delta G$ values of the BACE pairs is quite narrow and covers only 3.5 kcal/mol, while comparable in-size datasets cover a wider range of values of at least 5 kcal/mol. In effect inaccuracies of the method, as well as experimental measurements, get amplified.

\begin{table*}
\footnotesize
\centering
\begin{tabular}{|c|cc|cc|cc|cc|}
\hline
         &    ATM  &   &  FEP+\cite{wang2015accurate}    &    & Amber\cite{lee2020alchemical} &     &     pmx\cite{gapsys2020large} &  \\
         & r & MAE  & r & MAE   & r & MAE  & r & MAE  \\
\hline
\hline
MCL1     & \textbf{0.58$\pm$0.10} & 1.7$\pm$0.2 & 0.51$\pm$0.10 &  \textbf{1.4$\pm$0.2} & 0.51$\pm$0.10 & 1.3$\pm$0.2 & 0.32$\pm$0.11 & 1.2$\pm$0.2 \\
TYK2     & 0.63$\pm$0.17 & 0.9$\pm$0.2  & \textbf{0.70$\pm$0.15} & \textbf{0.7$\pm$0.2} & 0.58$\pm$0.17 & 0.9$\pm$0.2 & 0.64$\pm$0.16 & 1.0$\pm$0.2 \\
JNK1     & \textbf{0.69$\pm$0.13} & 0.6$\pm$0.1 & 0.59$\pm$0.15 & 0.8$\pm$0.1 & 0.59$\pm$0.15 & 0.7$\pm$0.2 & 0.65$\pm$0.14 & \textbf{0.5$\pm$0.1} \\
PTP1B    & 0.66$\pm$0.11 & 0.9$\pm$0.2 & 0.66$\pm$0.11 & 0.9$\pm$0.2 & 0.72$\pm$0.10 & 0.8$\pm$0.2 & \textbf{0.74$\pm$0.10} & \textbf{0.8$\pm$0.2} \\
CDK2     & 0.58$\pm$0.17 & 1.0$\pm$0.2 & 0.39$\pm$0.19 & 0.9$\pm$0.2 & 0.46$\pm$0.19 & 0.9$\pm$0.2 & \textbf{0.61$\pm$0.17} & \textbf{0.7$\pm$0.2} \\
Thrombin & \textbf{0.61$\pm$0.21} & 0.8$\pm$0.2 & 0.42$\pm$0.24 & 0.8$\pm$0.2 & 0.31$\pm$0.25 & \textbf{0.4$\pm$0.2} & -0.02$\pm$0.27 & 0.5$\pm$0.2\\
p38      & 0.71$\pm$0.09 & 1.0$\pm$0.2 & 0.79$\pm$0.08 & 0.8$\pm$0.2 & \textbf{0.80$\pm$0.08} & 0.8$\pm$0.2 & 0.78$\pm$0.08 & \textbf{0.7$\pm$0.2} \\
BACE     & 0.42$\pm$0.12 & 1.2$\pm$0.2  & \textbf{0.61$\pm$0.11} & \textbf{0.8$\pm$0.2} & 0.54$\pm$0.11 & 0.9$\pm$0.2 & 0.49$\pm$0.12 & 0.9$\pm$0.2  \\
\hline
ALL      & 0.59$\pm$0.11 & 1.0$\pm$0.1 & \textbf{0.60$\pm$0.11} & 0.9$\pm$0.1 & 0.58$\pm$0.11 & 0.9$\pm$0.1 & 0.56$\pm$0.11 & \textbf{0.8$\pm$0.1}\\
\hline
\end{tabular}
\caption{\label{tab:result_table} Comparison of the performance of free energy methods. Pearson correlation (r) and Mean Absolute Error (MAE) in kcal/mol for the 8 tested Protein Targets. }
\end{table*}

When considering absolute deviations from the experimental references, ATM displayed consistently poorer performance than the other methodologies. ATM's MAE metric is the highest among the three methods considered in the comparison. However, the difference between ATM and other methodologies is not very high in most cases. The exception to this is BACE, where, consistent with the previously mentioned results, the differences are the highest. 

\begin{figure}
\centering
\includegraphics[width=\columnwidth]{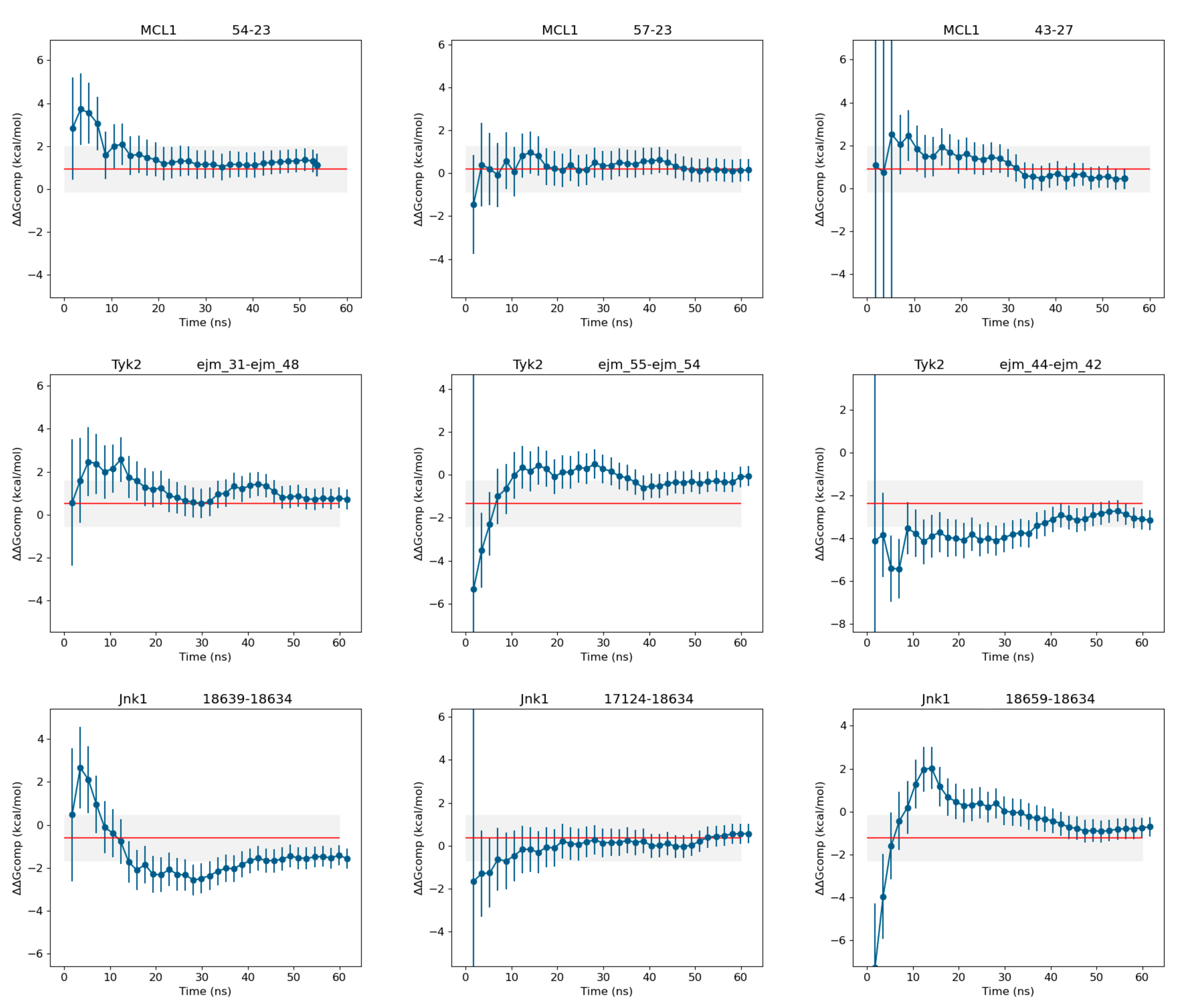}
\caption{\label{fig:convergence_plots} Free energy convergence as a function of time for a series of ligand pairs of MCL-1, TYK2 and JNK1. The red line corresponds to the experimental $\Delta \Delta G$ value. }
\end{figure}

In terms of convergence, we observed that 50 to 60 ns per $\Delta \Delta G$ estimate (2.3-2.8 ns per $\lambda$) tends to be sufficient. Convergence analysis over time shows good convergence for most cases as illustrated in Figure (Figure \ref{fig:convergence_plots}). The variance of the predicted $\Delta \Delta G$ tends to level off for a majority of cases around 50 ns per $\Delta \Delta G$ estimate. We also performed longer simulations for a series of ligand pairs to evaluate if that extended sampling time was causing any drift on the predicted $\Delta \Delta G$ (Supporting Figure \ref{fig:Converge_longer}) where we observed that obtained values were stable.  Furthermore, we analysed the perturbation energy distributions for every $\lambda$-state (Supporting Figures \ref{fig:perturbation_MCL1},\ref{fig:perturbation_TYK2},\ref{fig:perturbation_JNK1}). This analysis helps to determine if the system has converged or will converge in a reasonable amount of time. A poor overlap between perturbation energy distributions are indicative of unreliable relative free energy estimates that could require the design of alternative alchemical routes. We have observed that in some of the ligand pairs that showed poor correlation with both experimental and calculated values from other approaches, there tends to be a poor overlap between perturbation energy distributions of nearby $\lambda$-states, even after conducting multiple replicates. From these results we can say that we obtain an analogous convergence performing a similar sampling time than the other mentioned methodologies in this work. The issue of convergence of binding free energy calculations is a very complex topic that we intend to investigate in future work.

We have observed that ATM's performance metrics are significantly skewed by poor relative binding free energy predictions involving a relatively small number of problematic ligands (Supporting Figure \ref{fig:problematic_ligands}). These ligands might be affected by force field parameterization issues or some specific aspects of the ATM methodology. In terms of ligand force fields, OPLS3 was used for FEP+, GAFF2 was used in two of the approaches (ATM and Amber), and a consensus of GAFF2 with CGenFF in pmx. Given that all methods perform similarly, the accuracy of the forcefield is of the same order as the precision of the methods. In relation to this aspect, Merck published a series of guidelines for FEP calculations with the requirement of an RMSE  lower than 1.3 kcal/mol in the validation phase.\cite{schindler2020large}. As we can observe in Table \ref{tab:result_table} ATM fulfills this requirement for most of the analyzed systems in this study. 

One major difference between ATM and the other methods is that ATM models explicitly binding/unbinding processes in the alchemical space. For example, a flexible ligand with different conformational propensities when bound vs when in solution, will undergo an actual conformational transition. In double-decoupling instead, the transformation is applied to the bound and solution conformations individually and a conformational transition is not necessarily required to reach convergence. We believe that this is both a strength and a weakness of ATM. When conformational changes are important, especially if there are differences in the conformational transition between the ligands, ATM is expected to be superior to other methods as it explicitly models the transitions. When considering rigid ligands or ligands with similar transitions, the conformational rearrangements will cancel out and the extra work ATM needs to do is unnecessary and might hurt convergence. We intend to study these aspects in more detail in future work.

\section{Conclusion}

In this study, we evaluated the performance of the Alchemical Transfer Method (ATM), a novel approach for predicting protein-ligand binding affinities. We benchmarked it against the dataset of Wang \textit{et al.}\cite{wang2015accurate}, one of the most popular data sets on the evaluation of free binding energy methodologies. Our results showed that ATM is a competitive approach for predicting binding affinities, matching or even surpassing the performance of other state-of-the-art methods in terms of Pearson correlation. While mean absolute errors were slightly higher compared to other methods, ATM is a promising approach for the estimation of relative binding free energies.

Unlike other methods, ATM does not require splitting of binding free energy calculations into receptor and solvation legs or the use of softcore pair potentials. Furthermore, ATM is implemented in the open-source OpenMM MD engine, which is freely available. Its flexibility opens up the possibility for further improvement of the method through the use of new force fields, such as neural network potentials.

\section{Data and software availability}
The calculated free energy values, ligand and protein structures as well as preparation scripts are available at: \url{https://github.com/compsciencelab/ATM_benchmark}

\section{Acknowledgement}
This project has received funding from
the European Union’s Horizon 2020 research and innovation programme under grant agreement No. 823712;
and the project PID2020-116564GB-I00 has been funded by MCIN / AEI / 10.13039/501100011033; the Torres-Quevedo Programme from the Spanish National Agency for Research (PTQ2020-011145 / AEI / 10.13039/501100011033).
Research reported in this publication was supported by the National Institute of General Medical Sciences (NIGMS) of the National Institutes of Health under award number GM140090. {E. G.} acknowledges support from the United States' National Science Foundation (NSF CAREER 1750511).
The content is solely the responsibility of the authors and does not necessarily represent the official views of the National Institutes of Health.
\clearpage
\onecolumn

\section{Supporting Information} \label{sec:sup_methods}
\setcounter{figure}{0}  
\setcounter{table}{0}

\subsection*{Supporting Methods}

\begin{table}
\footnotesize
\centering
\begin{tabular}{c|c|c|c|c|c}
\hline
             $\lambda$   &  $\lambda_1$  & $\lambda_2$  & $\alpha$  &  $u_0$  &  $w_0$\\

\hline
 0.00  & 0.00 & 0.00 & 0.10 & 110 & 0 \\
 0.05  & 0.00 & 0.10 & 0.10 & 110 & 0 \\
 0.10  & 0.00 & 0.20 & 0.10 & 110 & 0 \\
 0.15  & 0.00 & 0.30 & 0.10 & 110 & 0 \\
 0.20  & 0.00 & 0.40 & 0.10 & 110 & 0 \\
 0.25  & 0.00 & 0.50 & 0.10 & 110 & 0 \\
 0.30  & 0.10 & 0.50 & 0.10 & 110 & 0 \\
 0.35  & 0.20 & 0.50 & 0.10 & 110 & 0 \\
 0.40  & 0.30 & 0.50 & 0.10 & 110 & 0 \\
 0.45  & 0.40 & 0.50 & 0.10 & 110 & 0 \\
 0.50  & 0.50 & 0.50 & 0.10 & 110 & 0 \\

\end{tabular}
\caption{\label{tab:params_table}Alchemical schedule of the Solftplus Alchemical Potential for the two legs for the alchemical transformations. $\alpha$ values are in (kcal/mol)$^{-1}$ and $u_0$ and $w_0$ are in kcal/mol }
\end{table}

\begin{figure}
\centering
\includegraphics[width=0.7\textwidth]{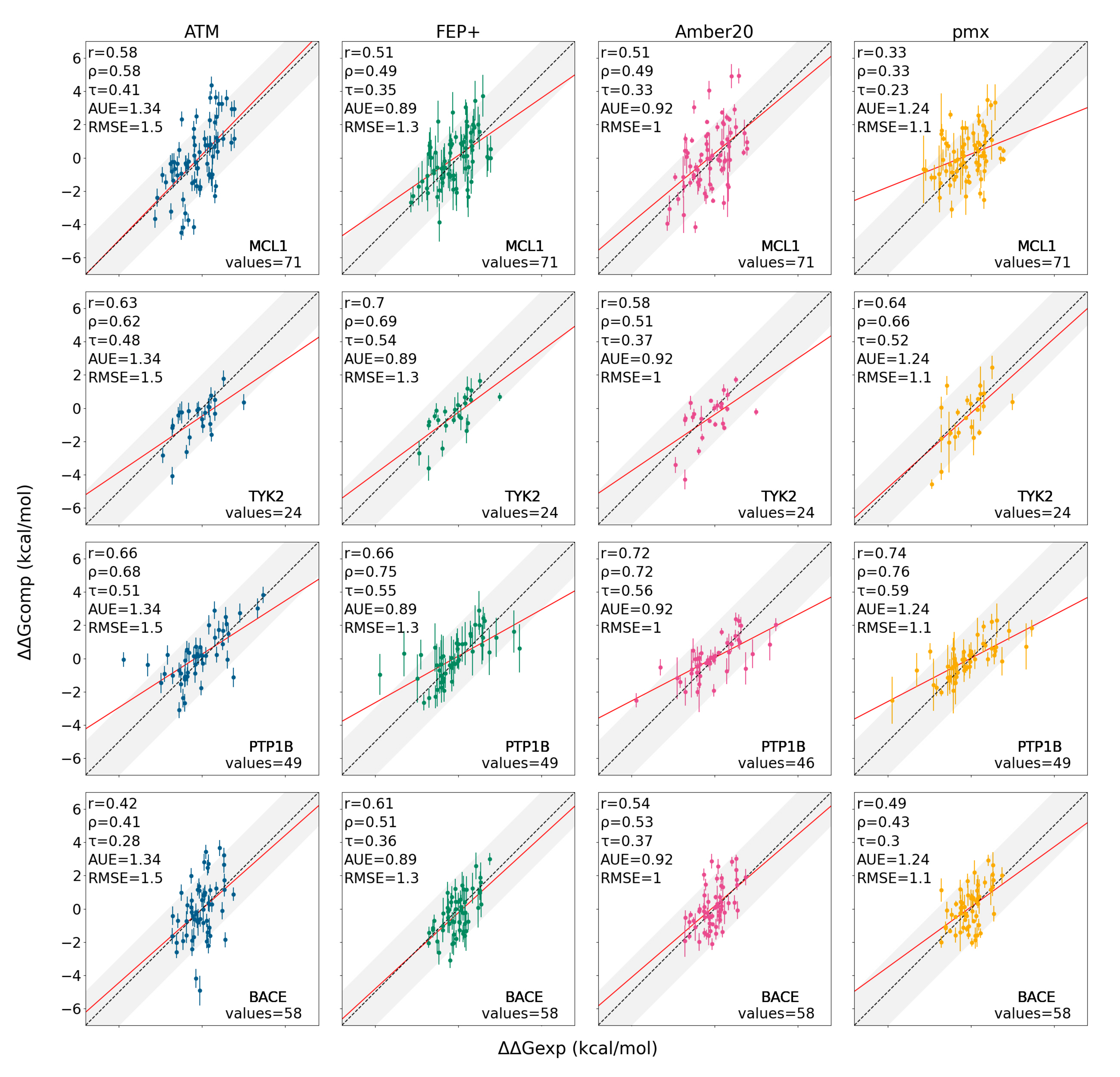}
\caption{\label{fig:comparison_corrsSI_1} Scatterplots for the calculated $\Delta\Delta$G estimated against the experimental measurements and compared to other methodologies for MCL-1, TYK2, PTP1B and BACE systems. The first column represents calculations performed with ATM. The other columns contain data from benchmark studies of FEP+\cite{wang2015accurate}, Amber\cite{lee2020alchemical} and pmx\cite{gapsys2020large}. }
\end{figure}

\begin{figure}
\centering
\includegraphics[width=0.7\textwidth]{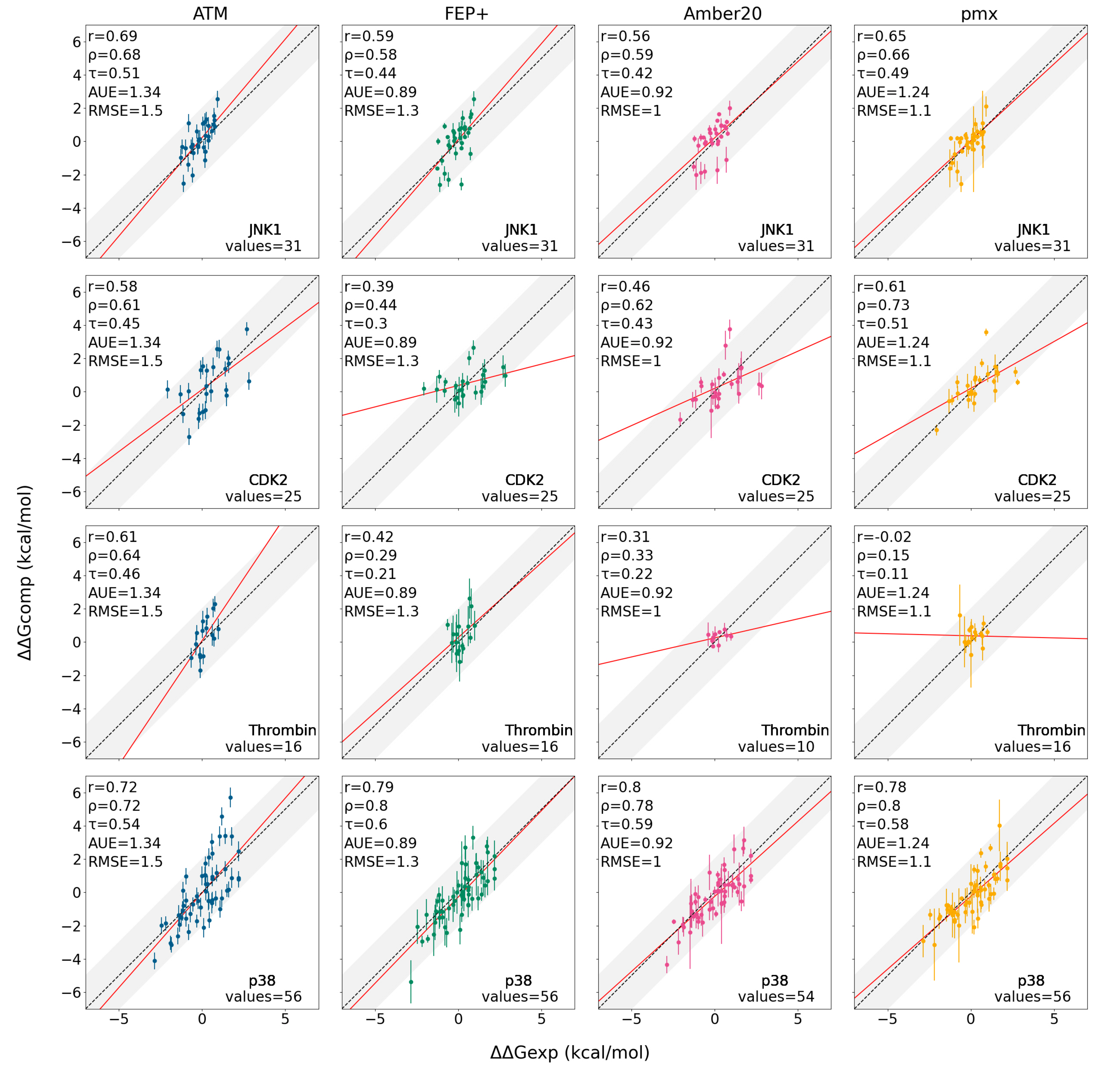}
\caption{\label{fig:comparison_corrsSI_2} Scatterplots for the calculated $\Delta\Delta$G estimated against the experimental measurements and compared to other methodologies for JNK1, CDK2, Thrombin and p38 systems. The first column represents calculations performed with ATM. The other columns contain data from benchmark studies of FEP+\cite{wang2015accurate}, Amber\cite{lee2020alchemical} and pmx\cite{gapsys2020large}. }
\end{figure}

\begin{figure}
\centering
\includegraphics[width=0.7\textwidth]{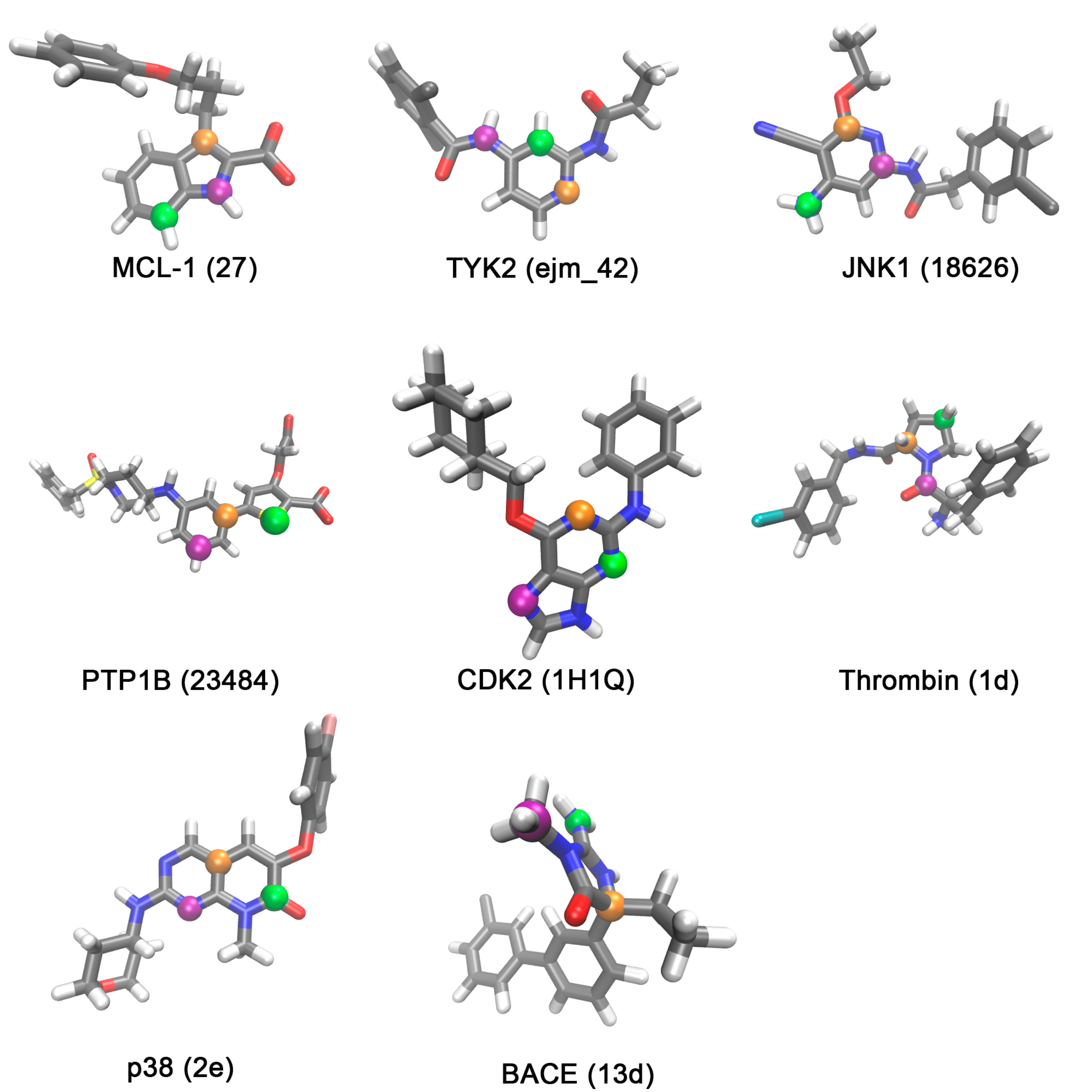}
\caption{\label{fig:ref_alignment} Reference atoms selected for the applied restraining potential in every system. selected atoms of each molecule define a cartesian coordinate system with the orange atom at the origin, a \textit{z} axis along the orange to green direction, and the purple atom oriented at the \textit{xz} plane}
\end{figure}

\begin{figure}
\centering
\includegraphics[width=\linewidth]{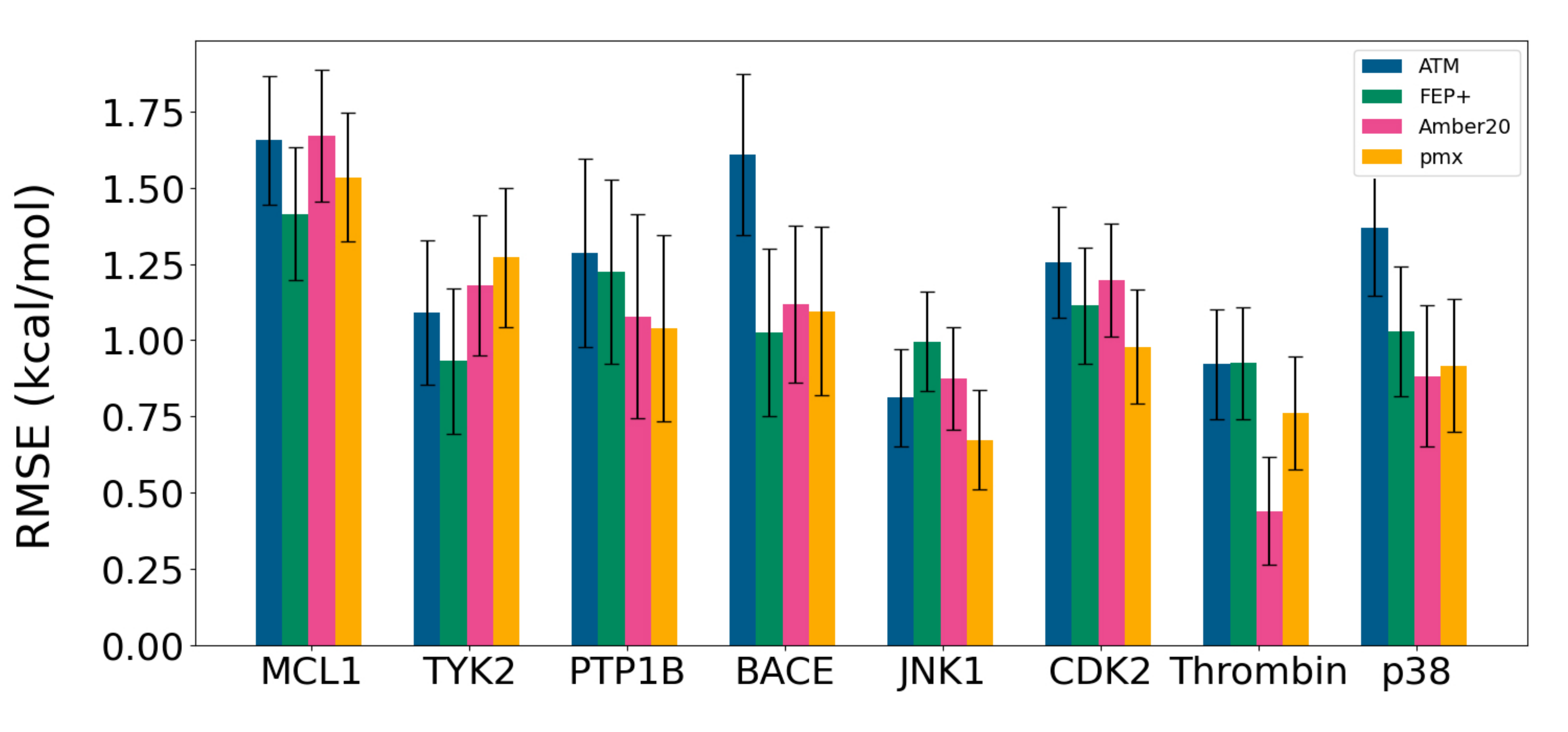}
\caption{\label{fig:comparison_RMSE} Root Mean Square Error (RMSE) for each protein-ligand system calculated with ATM and reported through other methodologies such as FEP+\cite{wang2015accurate}, Amber\cite{lee2020alchemical} and pmx\cite{gapsys2015pmx}.}
\end{figure}

\begin{figure}
\centering
\includegraphics[width=\linewidth]{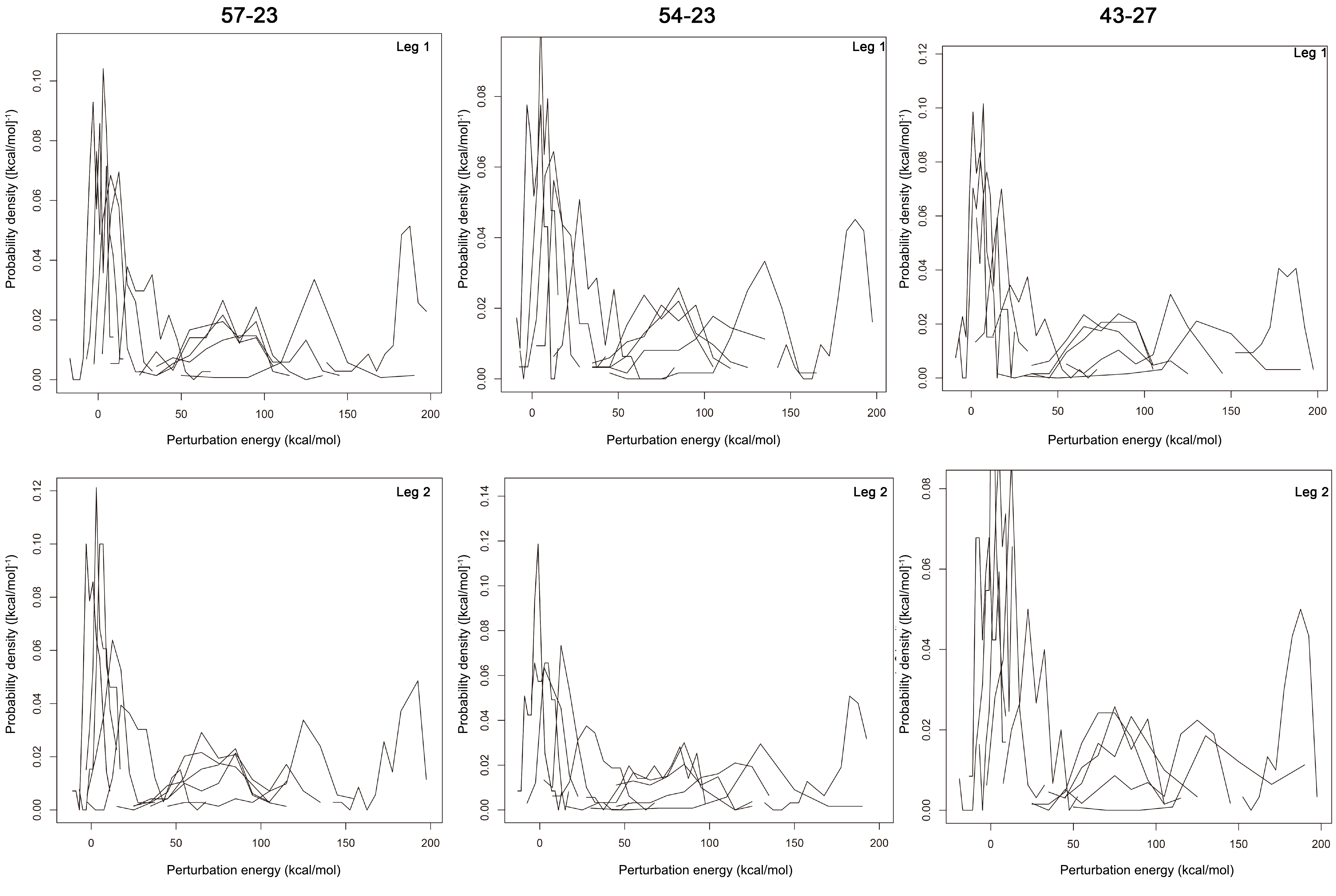}
\caption{\label{fig:perturbation_MCL1} Distribution of the perturbation energies for a series of ligand pairs coresponding to the MCL-1 system. Each line corresponds to a $\lambda$. Top row refers to the first leg an bottom row to the second one. }
\end{figure}

\begin{figure}
\centering
\includegraphics[width=\linewidth]{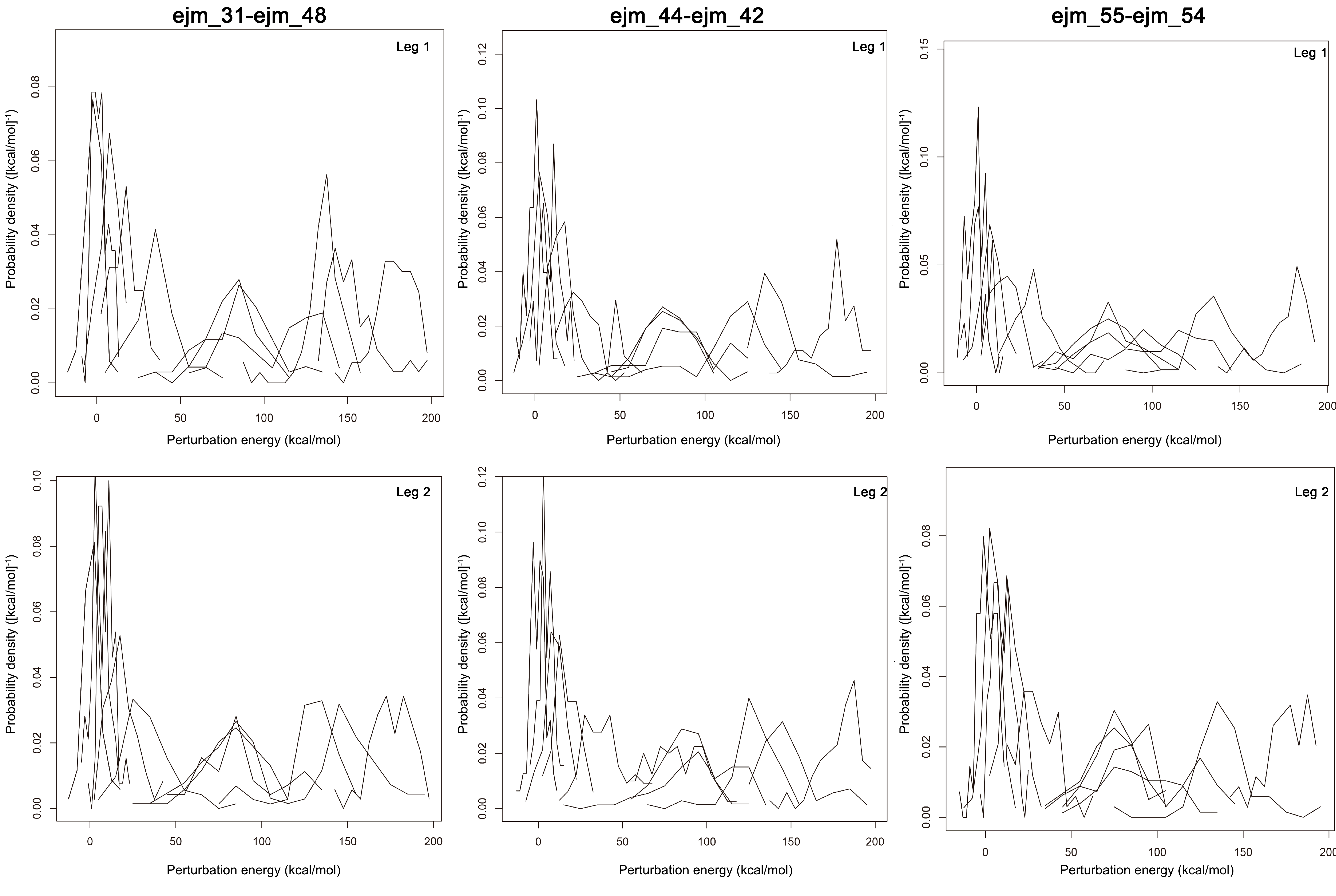}
\caption{\label{fig:perturbation_TYK2} Distribution of the perturbation energies for a series of ligand pairs coresponding to the TYK2 system. Each line corresponds to a $\lambda$. Top row refers to the first leg an bottom row to the second one. }
\end{figure}

\begin{figure}
\centering
\includegraphics[width=\linewidth]{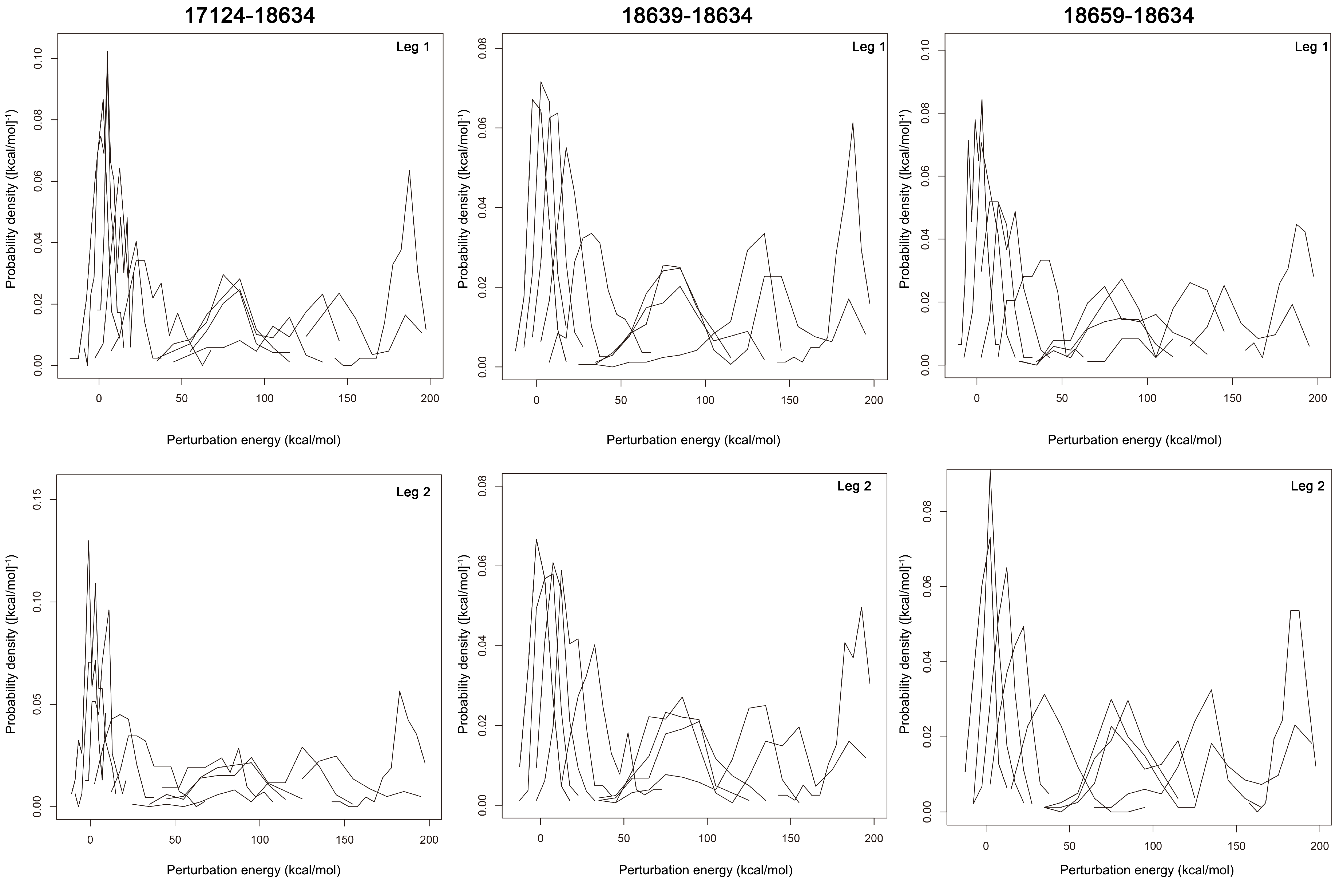}
\caption{\label{fig:perturbation_JNK1} Distribution of the perturbation energies for a series of ligand pairs coresponding to the JNK1 system. Each line corresponds to a $\lambda$. Top row refers to the first leg an bottom row to the second one.}
\end{figure}

\begin{figure}
\centering
\includegraphics[width=\linewidth]{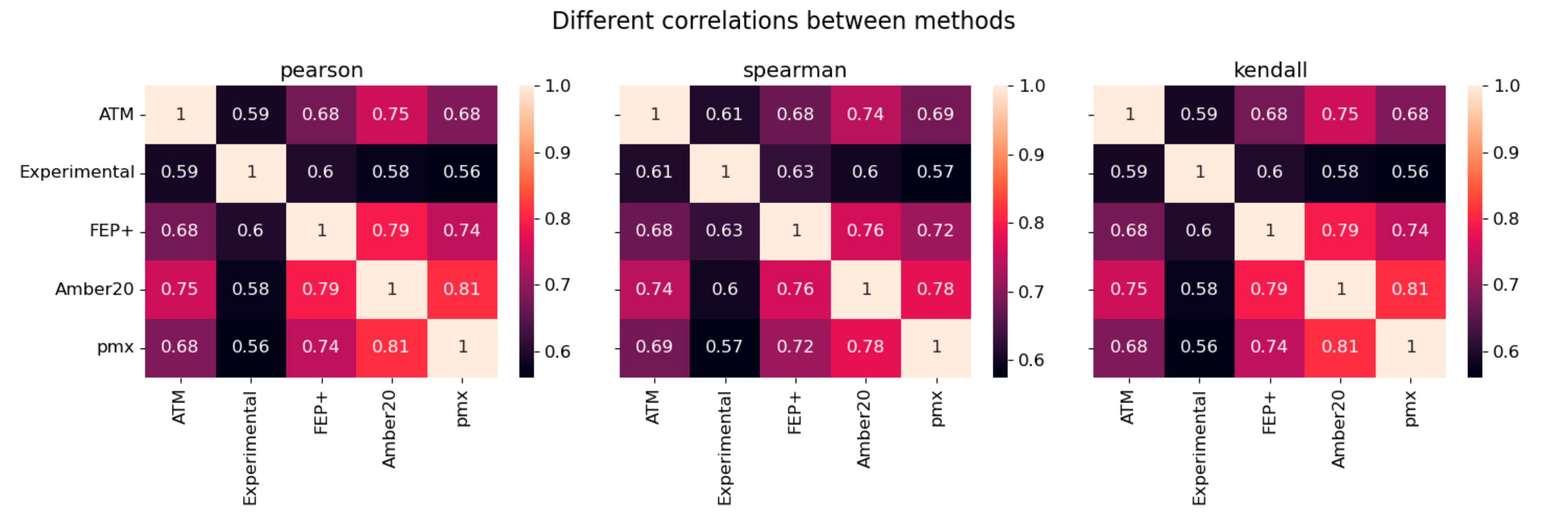}
\caption{\label{fig:correlation_methods} Pearson, spearman and kendall correlations for all data points of ATM against experimental data and the other compared methodologies: FEP+\cite{wang2015accurate}, Amber\cite{lee2020alchemical} and pmx\cite{gapsys2020large}}
\end{figure}

\begin{figure}
\centering
\includegraphics[width=\linewidth]{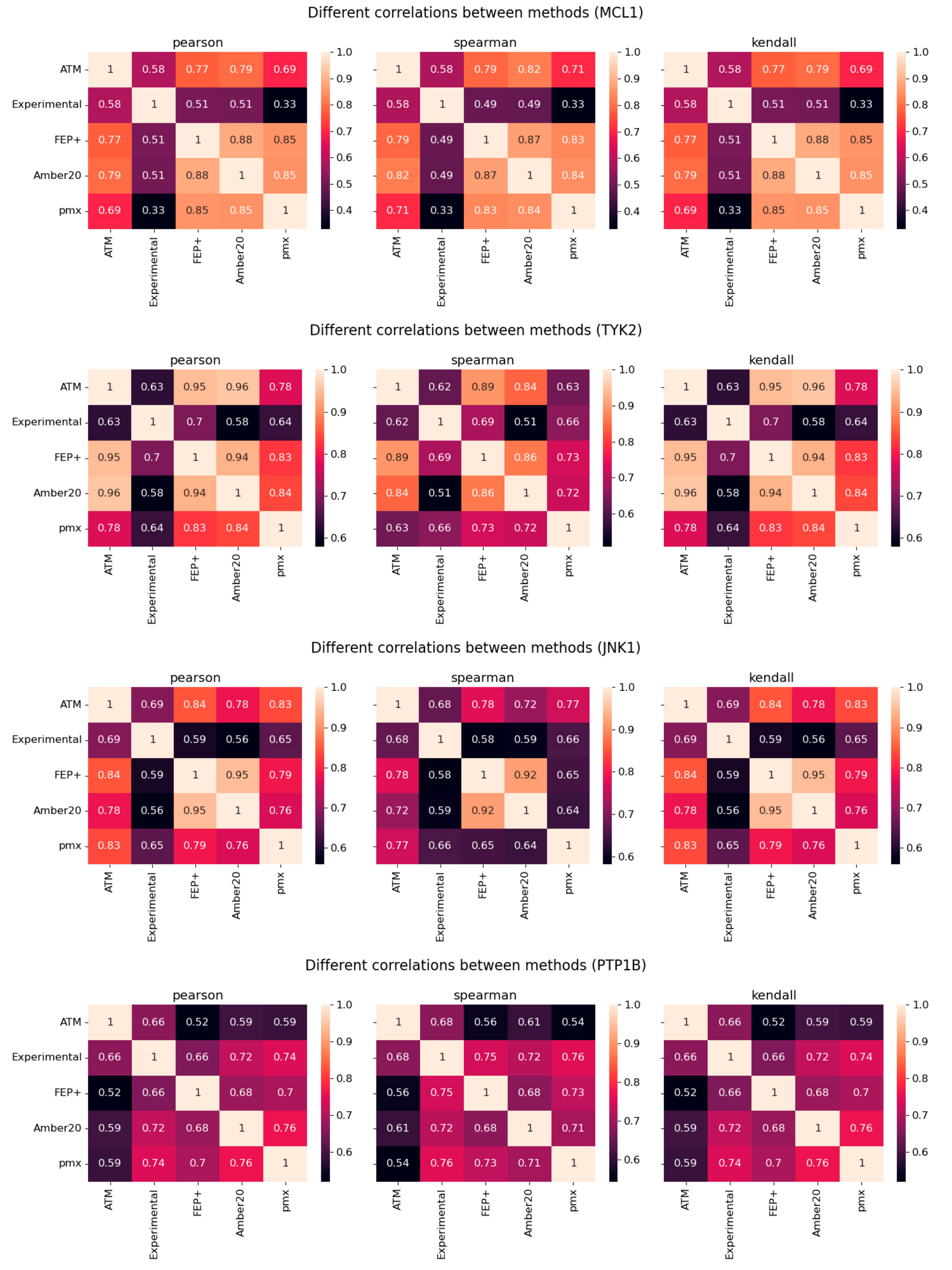}
\caption{\label{fig:correlation_methods_prot_1} Pearson, spearman and kendall correlations for the MCL1, TYK2, JNK1 and PTP1B systems of ATM against experimental data and the other compared methodologies: FEP+\cite{wang2015accurate}, Amber\cite{lee2020alchemical} and pmx\cite{gapsys2020large}}
\end{figure}

\begin{figure}
\centering
\includegraphics[width=\linewidth]{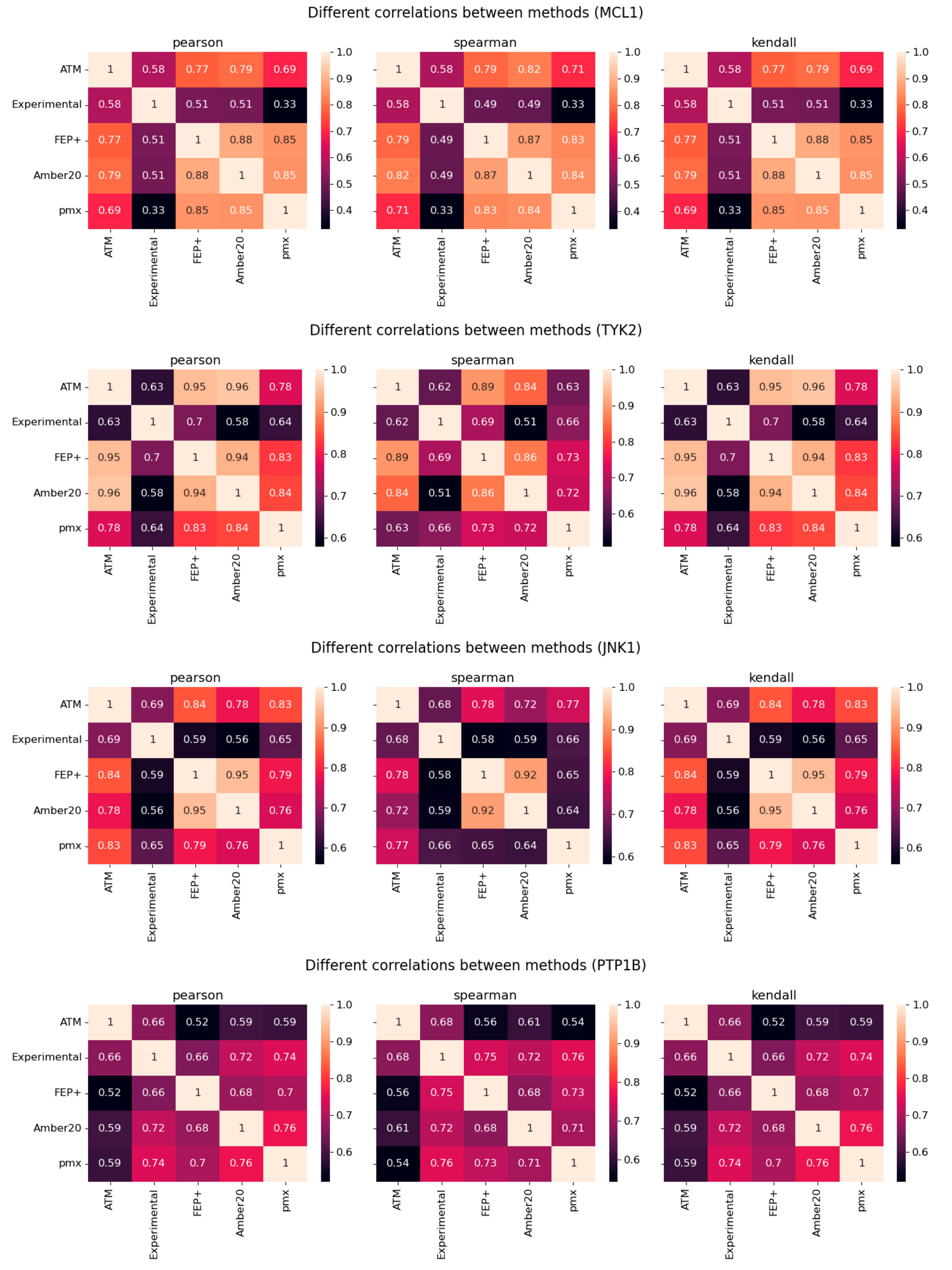}
\caption{\label{fig:correlation_methods_prot_2} Pearson, spearman and kendall correlations for the BACE, p38, CDK2 and Throbmin systems of ATM against experimental data and the other compared methodologies: FEP+\cite{wang2015accurate}, Amber\cite{lee2020alchemical} and pmx\cite{gapsys2020large}}
\end{figure}

\begin{figure}
\centering
\includegraphics[width=\linewidth]{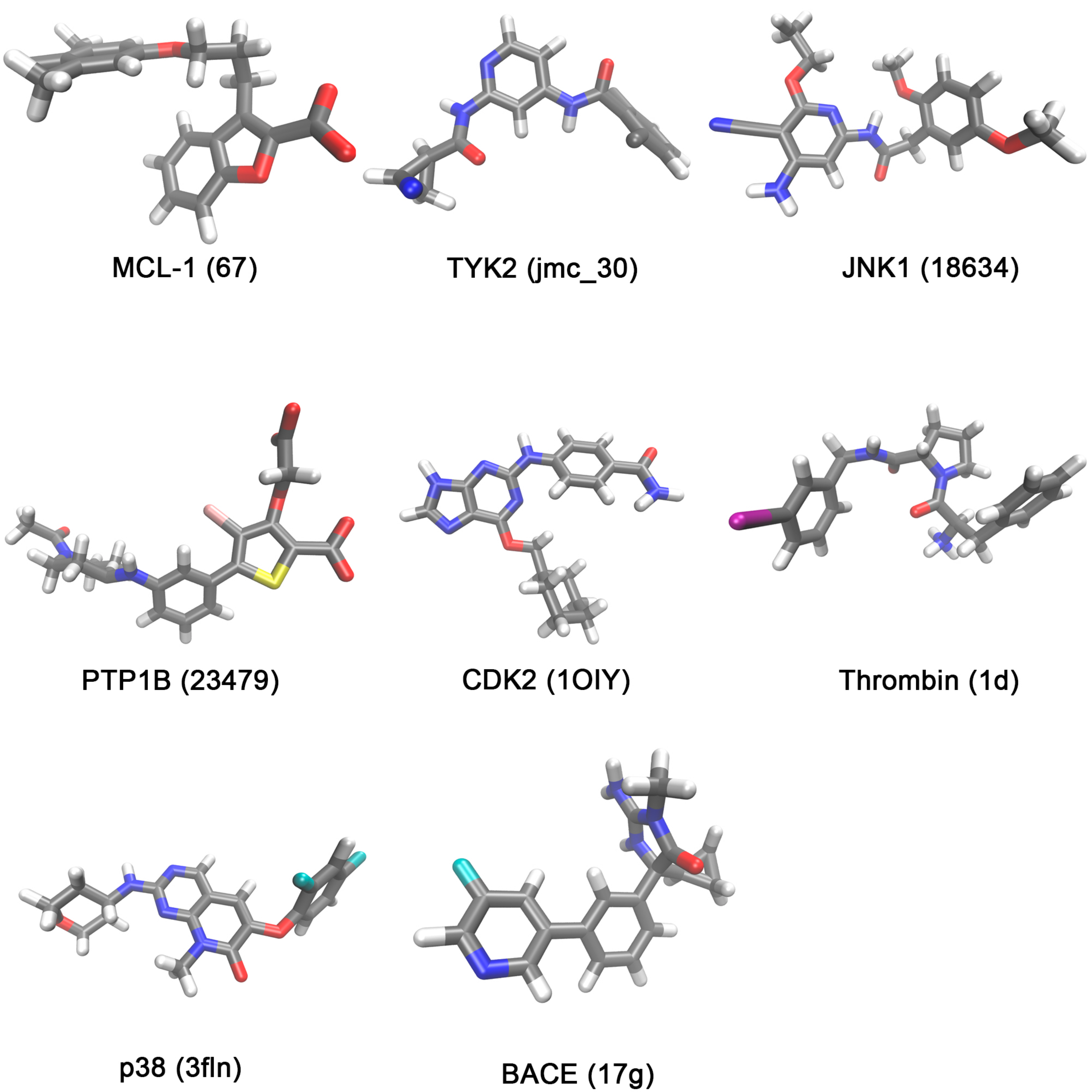}
\caption{\label{fig:problematic_ligands} Structures of some of the ligands that showed poor correlation with experimental values in two or more instances}
\end{figure}

\begin{figure}
\centering
\includegraphics[width=\linewidth]{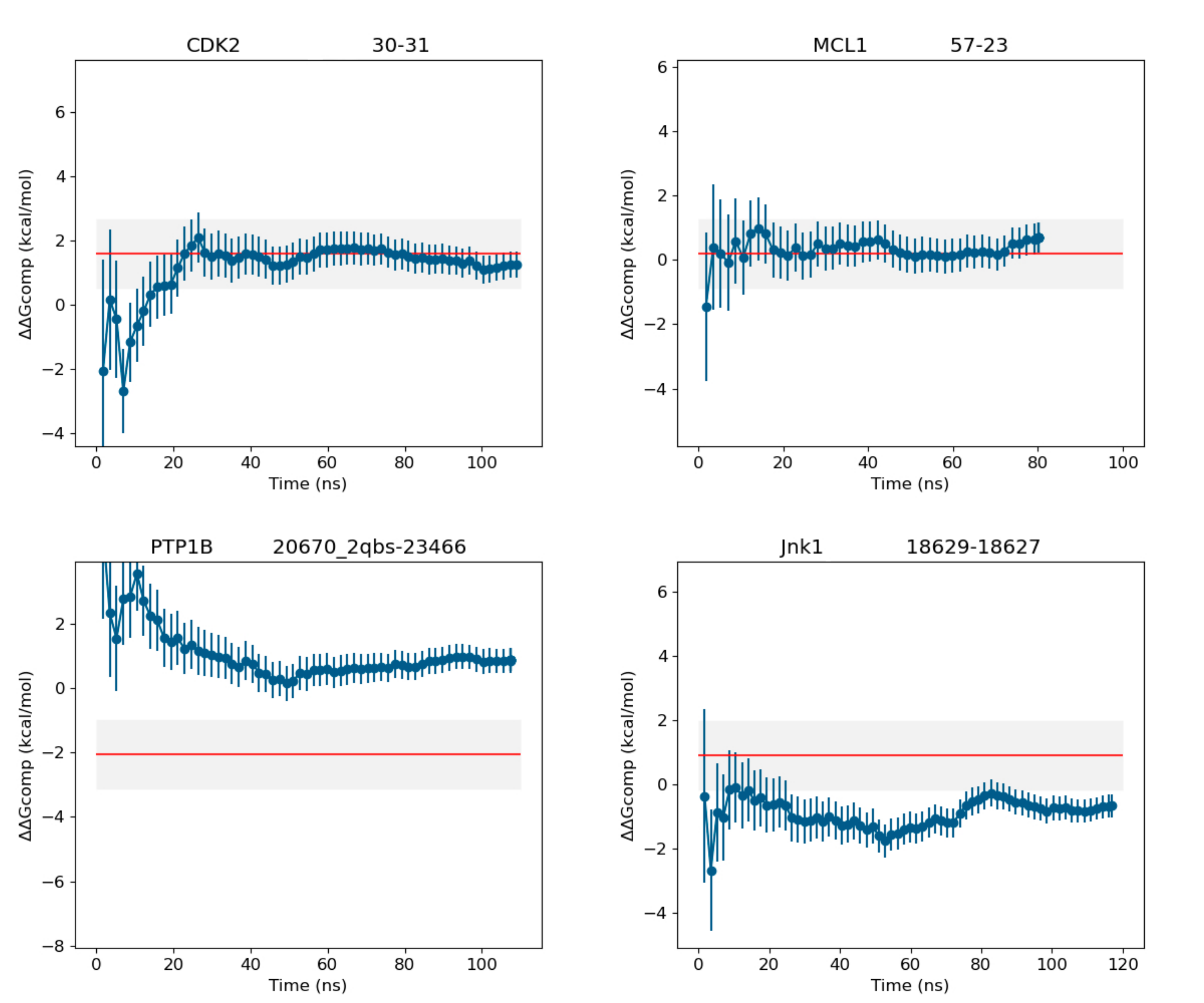}
\caption{\label{fig:Converge_longer}  Free energy convergence as a function of time for a series of ligand pairs of CDK2, MCL-1, PTP1B and JNK1 with a greater sampling than 50ns. The red line corresponds to the experimental $\Delta \Delta G$ value. }
\end{figure}

\bibliography{ref}

\end{document}